\documentstyle[aps,prb]{revtex}
\begin{document}
\title{Superfluid Helium 3:\\ Link between Condensed Matter Physics and Particle
Physics\footnote{To be published in Acta Physica Polonica B
[Proceedings of the XL Jubilee Cracow School of Theoretical
Physics on "Quantum Phase Transitions in High Energy and Condensed
Matter Physics"; 3-11 June, 2000, Zakopane, Poland]}}
\author{D. Vollhardt$^1$ and P.  W\"{o}lfle$^2$}
\address{$^1$Theoretical Physics III, Center for
Electronic Correlations and Magnetism,\\ University of Augsburg,
86135 Augsburg, Germany\\ $^2$Institut f\"{u}r Theorie der
Kondensierten Materie, Universit\"{a}t Karlsruhe, D-76128
Karlsruhe, Germany }
\date{\today}
\maketitle

\begin{abstract}
The discovery of the superfluid phases of Helium 3 in 1971 opened
the door to one of the most fascinating systems known in condensed
matter physics. Superfluidity of Helium 3, originating from pair
condensation of Helium 3 atoms, turned out to be the ideal
testground for many fundamental concepts of modern physics, such
as macroscopic quantum phenomena, (gauge-)symmetries and their
spontaneous breakdown, topological defects, etc.\ Thereby the
superfluid phases of Helium 3 enriched condensed matter physics
enormously. In particular, they contributed significantly -- and
continue to do so -- to our understanding of various other
physical systems, from heavy fermion and high-$T_{c}$
superconductors all the way to neutron stars, particle physics,
gravity and the early universe. A simple introduction into the
basic concepts and questions is presented.


\end{abstract}

\section{The quantum liquids $^{3}$He and $^{4}$He}

There are two stable isotopes of the chemical element Helium: Helium 3 and
Helium 4, conventionally denoted by $^{3}$He and $^{4}$He, respectively.
From a microscopic point of view, Helium atoms are structureless, spherical
particles interacting via a two-body potential that is well understood. The
attractive part of the potential, arising from weak van der Waals-type
dipole (and higher multipole) forces, causes Helium gas to condense into a
liquid state at temperatures of 3.2 K and 4.2 K for $^{3}$He and $^{4}$He,
respectively, at normal pressure. The pressure versus temperature phase
diagrams of $^{3}$He and $^{4}$He are shown in Fig. 1. When the temperature
is decreased even further one finds that the Helium liquids, unlike all
other liquids, do not solidify unless a pressure of around 30 bar is
applied. This is the first remarkable indication of macroscopic quantum
effects in these systems. The origin of this unusual behaviour lies in the
quantum-mechanical uncertainty principle, which requires that a quantum
particle can never be completely at rest at given position, but rather
performs a zero-point motion about the average position. The smaller the
mass of the particle and the weaker the binding force, the stronger these
oscillations are. In most solids the zero-point motion is confined to a
small volume of only a fraction of the lattice-cell volume. In the case of
Helium, however, two features combine to prevent the formation of a
crystalline solid with a rigid lattice structure: (i) the strong zero-point
motion arising from the small atomic mass (Helium is the second-lightest
element in the periodic table); and (ii) the weakness of the attractive
interaction due to the high symmetry of these simple atoms. It is this very
property of Helium -- of staying liquid -- that makes it such a valuable
system for observing quantum behavior on a macroscopic scale. Quantum
effects are also responsible for the strikingly different behaviors of $^{4}$%
He and $^{3}$He at even lower temperatures. Whereas $^{4}$He undergoes a
second-order phase transition into a state later shown to be superfluid,
i.e. where the liquid is capable of flowing through narrow capillaries or
tiny pores without friction, no such transition is observed in liquid $^{3}$%
He in the same temperature range (Fig. 1). The properties of liquid $^{3}$He
below 1 K are nevertheless found to be increasingly different from those of
a classical liquid. It is only at a temperature roughly one thousandth of
the transition temperature of $^{4}$He that $^{3}$He also becomes
superfluid, and in fact forms ${\it several}$ superfluid phases, each of
which has a much richer structure than that of superfluid $^{4}$He.

The striking difference in the behaviors of $^{3}$He and $^{4}$He at low
temperatures is a consequence of the laws of quantum theory as applied to
systems of identical particles, i.e. the laws of quantum statistics. The $%
^{4}$He atom, being composed of an even number of electrons and nucleons,
has spin zero and consequently obeys Bose-Einstein statistics. In contrast,
the $^{3}$He nucleus consists of ${\em {three}}$ nucleons, whose spins add
up to give a total nuclear spin of $I=\frac{1}{2}$, making the total spin of
the entire $^{3}$He atom $\frac{1}{2}$ as well. Consequently liquid $^{3}$He
obeys Fermi-Dirac statistics. So it is the tiny nuclear spin, buried deep
inside the Helium atom, that is responsible for all the differences of the
macroscopic properties of the two isotopes.

Since in a Bose system single-particle states may be multiply
occupied, at low temperatures this system has a tendency to
condense into the lowest-energy single-particle state
(Bose-Einstein condensation). It is believed that the superfluid
transition in $^{4}$He is a manifestation for Bose-Einstein
condensation. The all-important qualitative feature of the Bose
condensate is its phase rigidity, i.e. the fact that it is
energetically favourable for the particles to condense into a
single-particle state of fixed quantum-mechanical phase, such that
the global gauge symmetry spontaneously broken. As a consequence,
macroscopic flow of the condensate is (meta-)stable, giving rise
to the phenomenon of superfluidity. In a Fermi system, on the
other hand, the Pauli exclusion principle allows only single
occupation of fermion states. The ground state of the Fermi gas is
therefore the one in which all single-particle states are filled
up to a limiting energy, the Fermi energy $E_{F}$. As predicted by
Landau\cite{Landau} and later verified experimentally (for a
review see \cite{Wheatley66}), the properties of $^{3}$He well
below its Fermi temperature ${{T}_{F}={E}_{F}/{k}_{B}{\approx
}1\mbox{K}}$ are similar to those of a degenerate Fermi gas. In
particular the formation of a phase-rigid condensate is not
possible in this framework. Until the mid-1950s a superfluid phase
of liquid $^{3}$He was therefore believed to be ruled out. On the
other hand, it is most remarkable that the property of
superfluidity \cite{London50,London54} was indeed first discovered
experimentally in a ${\it Fermi}$ system, namely that of the
``liquid'' of conduction electrons in a superconducting metal
\cite{Kamerlingh11}. The superfluidity of $^{4}$He was only found
more than 25 years later. \newline

\section{Pair condensation in a Fermi liquid}

As demonstrated by Bardeen, Cooper and Schrieffer (BCS) in 1957
\cite{BCS} the key to the theory of superconductivity is the
formation of ``Cooper pairs'', i.e. pairs of electrons with
opposite momentum ${\bf k}$ and spin projection $\sigma $: (${\bf
k}\uparrow ,-{\bf k}\downarrow $). In the case of conventional
superconductors the Cooper pairs are structureless objects, i.e.
the two partners form a spin-singlet state in a relative s-wave
orbital state. These Cooper pairs have total spin zero and may
therefore be looked upon in a way as composite bosons, which all
have the same pair wave function and are all in the same
quantum-mechanical state. In this picture the transition to the
superconducting state corresponds to the Bose-condensation of
Cooper pairs, the condensate being characterized by macroscopic
quantum coherence. The concept of Bose-Einstein condensation is
appealing since key features of superconductivity like the
Meissner effect, flux quantization and superfluid mass currents in
conventional superconductors are naturally implied. Nevertheless,
since the theory of conventional superconductivity is firmly based
on BCS theory, the concept of a Bose-Einstein condensation of
Cooper pairs traditionally did not receive much attention (or was
even considered to be downright wrong). Within the context of
superfluid $^{3}$He, this notion was taken up again by Leggett
\cite{Leggett80a,Leggett80b}, who argued that tightly bound
Bose-Einstein-condensed molecules on the one hand and Cooper pairs
on the other may be viewed as extreme limits of the same
phenomenon. This approach, which was quite provocative at the
time, is now well accepted \cite
{Zwerger92,Nozieres95,Randeria95}. However, the original idea that
at T$_{c}$ Cooper pairs form and automatically Bose-condense has
been revised since then. Apparently Cooper pair formation is not a
separate phase transition but is rather a matter of thermal
equilibrium: for any finite coupling there exists a finite density
of pairs even above T$_{c}$, although in conventional
superconductors -- and even in high T$_{c}$ materials -- their
density is negligibly small. At weak coupling (BCS limit) the
condensation temperature and the (not well-defined) temperature of
pair formation practically coincide; they become different only at
very strong coupling (Bose limit). Similar ideas are also implicit
in several theoretical approaches to high-$T_{c}$
superconductivity.

While in free space an attractive force has to be sufficiently strong to
bind two electrons, inside the metal the presence of the filled Fermi sea of
conduction electrons blocks the decay of a Cooper pair, so that an ${\it %
arbitrarily}$ small attractive interaction leads to the formation of stable
Coopers pairs. The attractive interaction between the electrons of a Cooper
pair in a conventional superconducting metal is due to the exchange of
virtual phonons (electron-phonon interaction). If the phonon-mediated
interaction is strong enough to overcome the repulsive Coulomb interactions
between the two electrons then a transition into a superconducting state may
occur. On the other hand, any other mechanism leading to attraction between
electrons at the Fermi surface is equally well suited for producing
superconductivity.

Given the success of the BCS theory in the case of superconductivity, it was
natural to ask whether a similar mechanism might also work for liquid $^{3}$%
He. Since there is no underlying crystal lattice in the liquid that could
mediate the attractive force, the attraction must clearly be an intrinsic
property of the one-component $^{3}$He liquid itself. The main feature of
the interatomic $^{3}$He potential is the strong repulsive component at
short distances, and the weak van der Waals attraction at medium and long
distances. It soon became clear that, in order to avoid the hard repulsive
core and thus make optimal use of the attractive part of the potential, the $%
^{3}$He atoms would have to form Cooper-pairs in a state of ${\it nonzero}$
relative angular momentum $l$. In this case the Cooper-pair wave function
vanishes at zero relative distance, thus cutting out the most strongly
repulsive part of the potential. In a complementary classical picture one
might imagine the partners of a Cooper pair revolving about their centre of
gravity, thus being kept away from each other by the centrifugal force.

When the superfluid phases of $^{3}$He were finally discovered in 1971 at
temperatures of about 2.6 mK and 1.8 mK, respectively, by Osheroff,
Richardson and Lee \cite{Osheroff72a}, in an experiment actually designed to
observe a magnetic phase transition in solid $^{3}$He, the results came as a
great surprise.

\section{The superfluid phases of $^{3}$He}

Soon after the discovery of the phase transitions by Osheroff et
al. \cite {Osheroff72a} it was possible to identify altogether
${\it three}$ distinct stable superfluid phases of bulk $^{3}$He ;
these are referred to as the A, B and A$_{1}$ phases. In zero
magnetic field only the A and B phases are stable. In particular,
in zero field the A phase only exists within a finite range of
temperatures, above a critical pressure of about 21 bar. Hence its
region of stability in the pressure-temperature phase diagram has
a roughly triangular shape as shown in Fig. 1a. The B phase, on
the other hand, occupies the largest part of this phase diagram
and is found to be stable down to the lowest temperatures attained
so far. Application of an external magnetic field has a strong
influence on this phase diagram. First of all, the A phase is now
stabilized down to zero pressure. Secondly, an entirely new phase,
the A$_{1}$ phase, appears as a narrow wedge between the normal
state and the A and B phases.

Owing to the theoretical work on anisotropic superfluidity that had been
carried out before the actual discovery of superfluid $^{3}$He, progress in
understanding the detailed nature of the phases was very rapid. This was
clearly also due to the excellent contact between experimentalists and
theorists, which greatly helped to develop the right ideas at the right
time; see the reviews by Leggett \cite{Leggett75}, Wheatley \cite{Wheatley75}%
, and Lee and Richardson \cite{Lee78}. A comprehensive discussion
of the superfluid phases of $^{3}$He with a very extensive list of
references can be found in our book \cite{Vollhardt90} which also
provides most of the material presented in this article.

In particular, it fairly soon became possible to identify the A phase and
the B phase as realizations of the states studied previously by Anderson and
Morel \cite{Anderson60} and Balian and Werthamer \cite{Balian63},
respectively. Therefore the A phase is described by the so-called
``Anderson-Brinkman-Morel'' (ABM) state, while the B phase is described by
the ``Balian-Werthamer'' (BW) state. Consequently, ``A phase'' and ``ABM
state'' are now used as synonyms; the same is true in the case of ``B
phase'' and ``BW state''.

Although the three superfluid phases all have very different
properties, they have one important thing in common: the Cooper
pairs in all three phases are in a state with ${\it parallel}$
spin (S = 1) and relative orbital angular momentum $l=1$. This
kind of pairing is referred to as ``spin-triplet p-wave pairing''.
In contrast, prior to the discovery of the superfluid phases of
$^{3}$He, Cooper pairing in superconductors was only known to
occur in a state with opposite spins (S = 0) and $l=1$, i.e. in a
``spin-singlet s-wave state''. It should be noted that Cooper
pairs in a superconductor and in superfluid $^{3}$He are therefore
very different entities: in the former case pairs are formed by
point-like, structureless electrons and are spherically symmetric,
while in the case of $^{3}$He Cooper pairs are made of actual
atoms (or rather of quasiparticles involving $^{3}$He atoms) and
have an internal structure themselves.

\subsection{The Cooper pair wave function}

Quantum-mechanically, a spin-triplet configuration (S = 1) of two particles
has three substates with different spin projection $S_{z}$. They may be
represented as $\mid \uparrow \uparrow \rangle $ with $S_{z}=+1,2^{-1/2}(%
\mid \uparrow \downarrow \rangle +\mid \downarrow \uparrow )$ with $S_{z}=0$%
, and $\mid \downarrow \downarrow \rangle $ with $S_{z}=-1$. The pair wave
function $\Psi $ is in general a linear superposition of all three
substates, i.e.
\begin{equation}
\Psi =\psi _{1,+}({\bf k})\mid \uparrow \uparrow \rangle +\psi _{1,0}({\bf k}%
)(\mid \uparrow \downarrow \rangle +\mid \downarrow \uparrow \rangle )+\psi
_{1,-}({\bf k})\mid \downarrow \downarrow \rangle
\end{equation}
where $\psi _{1,+}({\bf k}),\psi _{1,0}({\bf k})$ and $\psi _{1,-}({\bf k})$
are the three complex-valued amplitudes of the respective substates. In the
case of a superconductor, where S = 0 and $l=0$, the pair wave function is
much simpler, i.e. it is given by only a single component
\begin{equation}
\Psi _{sc}=\psi _{0}(\mid \uparrow \downarrow \rangle -\mid \downarrow
\uparrow \rangle )
\end{equation}
with a single amplitude $\psi _{0}$.

So far we have only taken into account that, since S = 1, there are three
substates for the spin. The same is of course true for the relative orbital
angular momentum $l=1$ of the Cooper pair, which also has three substates $%
l_{z}=0,\pm 1$. This fact is important if we want to investigate
the amplitudes $\psi _{1,+}({\bf k})$ etc. further. They still
contain the complete information about the space (or momentum)
dependence of $\Psi $. The pair wave function $\Psi $ is therefore
characterized by three spin substates and three orbital substates,
i.e. by altogether 3$\times $3 = 9 substates with respect to the
spin and orbital dependence. Each of these nine substates is
connected with a complex-valued parameter. Here we see the
essential difference betweeen Cooper pairs with $S=l=0$
(conventional superconductors) and $S=l=1$($^{3}$He): their pair
wave functions are very different. In the former case a single
complex-valued parameter is sufficient for its specification, in
the latter case of superfluid $^{3}$He {\it nine} such parameters
are required. This also expresses the fact that a Cooper pair in
superfluid $^{3}$He has an internal structure, while that for a
conventional superconductor does not: because $l=1$, it is
intrinsically {\it anisotropic}. This anisotropy may conveniently
be described by specifying some direction with respect to a
quantization axis both for the spin and the orbital component of
the wave function.

In order to understand the novel properties of superfluid $^{3}$He, it is
therefore important to keep in mind that there are two characteristic
directions that specify a Cooper pair. Here lies the substantial difference
from a superconductor and the origin of the multitude of unusual phenomena
occurring in superfluid $^{3}$He: the structure of the Cooper pair is
characterized by {\it internal degrees of freedom}. Nevertheless, in both
cases the superfluid/superconducting state can be viewed as the condensation
of a macroscopic number of these Cooper pairs into the same
quantum-mechanical state, similar to a Bose-Einstein condensation, as
discussed above.

\section{Order parameter and broken symmetry}

\label{brokensymm}

In the normal liquid state the Cooper pair expectation value is
zero. Obviously in the superfluid a new state of order appears,
which spontaneously sets in at the critical temperature $T_{c}$.
This particular transition from the normal fluid to the
superfluid, i.e. into the ordered state, is called ``continuous'',
since the condensate -- and hence the state of order -- builds up
continuously. This fact may be expressed quantitatively by
introducing an ``order-parameter'' that is finite for $T<T_{c}$
and zero for $T\geq T_{c}$.

{\bf Ferromagnets: }A well-known example of such a transition is that from a
paramagnetic to a ferromagnetic state of a metal when the system is cooled
below the Curie temperature. In the paramagnetic regime the spins of the
particles are disordered such that the average magnetization $\langle {\bf M}%
\rangle $ of the system is zero. Mathematically speaking, the magnetization $%
{\bf M}$ is a real three-component vector. Invariance in the paramagnetic
phase under arbitrary global rotations in the three-dimensional spin space
implies that the spin system is invariant under a symmetry group $G=SO(3)$.
(For an introduction into the mathematical description of broken symmetries
by group theory see Ref. \cite{Vollhardt90}). By contrast, in the
ferromagnetic phase the spins are more or less aligned and $\langle {\bf M}%
\rangle $ is thus finite. In this case the system exhibits long-range order
of the spins. Clearly, the existence of a preferred direction ${\bf M}$ of
the spins implies that the symmetry of the ferromagnet under spin rotations
is reduced (``broken'') when compared with the paramagnet: the directions of
the spins are no longer isotropically distributed, and the system will
therefore no longer be invariant under the full spin rotation group $G=SO(3)$%
. So, in the ferromagnetic phase the system is only invariant under
rotations {\em about} the axis ${\bf M}$, corresponding to the group of
rotations $H=U(1)$ -- a subgroup of $SO(3)$; $H$ is called the {\em %
remaining symmetry} in the ferromagnetic phase. This phenomenon is called
``spontaneously broken symmetry''; it is of fundamental importance in the
theory of phase transitions. It describes the property of a macroscopic
system (i.e., a system in the thermodynamic limit) that is in a state that
does not have the full symmetry of the microscopic dynamics. The degree of
ordering is quantified by $|\langle {\bf M}\rangle |$, the magnitude of the
magnetization. Hence $|\langle {\bf M}\rangle |$ is called the ``order
parameter'' of the ferromagnetic state.

{\bf BCS superconductors and superfluid }$^{4}${\bf He: }The concept of
spontaneously broken symmetry also applies to superconductors and
superfluids. Here the order parameter measures the existence of Cooper pairs
and is given by the probability amplitude for a pair to exist at a given
temperature. In the case of conventional superconductors and superfluid $%
^{4} $He the order parameter is given by a single {\em complex} parameter $%
\Delta =\Delta _{0}e^{i\phi }$, with amplitude $\Delta _{0}$\ and
phase (''gauge'') $\phi $. Above $T_{c}$, in the normal phase, the
system is invariant under an arbitrary change of the phase $\phi
\rightarrow \phi ^{\prime }$, i.e. under a gauge transformation.
This gauge symmetry is due to the fact that the interparticle
forces conserve particle number, and is equivalent to the symmetry
group $G=U(1)$. Below $T_{c}$, when Cooper pairs have a finite
expectation value such that particle number is no longer
conserved, a particular value of $\phi $ is spontaneously preferred and the $%
U(1)$ symmetry is completely broken.

Gauge symmetry is spontaneously broken in any superfluid or superconductor.
In addition, in an odd-parity pairing superfluid, as in the case of $^{3}$%
He, where $l=1$, the pairs are necessarily in a spin-triplet
state, implying that rotational symmetry in spin space is broken,
just as in a magnet. At the same time, the anisotropy of the
Cooper-pair wave function in orbital space calls for a spontaneous
breakdown of orbital rotation symmetry, as in liquid crystals. All
three symmetries are therefore simultaneously broken in superfluid
$^{3}$He. What then are the order parameter and the spontaneously
broken symmetries of superfluid $^{3}$He?

\subsection{The order parameter and symmetry group describing superfluid $^{%
{\bf 3}}$He}

Owing to the fact that the Cooper pair wave function of superfluid $^{3}$He
is complex and has a spin-triplet ($S=1$), p-wave ($l=1$) structure with 3$%
\times $3 = 9 substates the order parameter describing the superfluid phase
may be written as a 3$\times $3 matrix $A_{\mu j}$ with complex components.
Here $\mu $=1,2,3 refers to spin space and $j$=1,2,3 to orbital (i.e. ${\bf k%
}$-) space. Obviously this order parameter has a considerably more
complicated structure than conventional order parameters.

If we only consider the interactions responsible for the formation
of the condensed state then the free energy of the system has to
be invariant under separate three-dimensional rotations in spin
space, in orbital space and under a gauge transformation. Hence
the symmetry group that allows for these symmetries is given by a
product of three independent symmetries \cite {Leggett75}

\begin{equation}
\ G=SO(3)_{L}\times SO(3)_{S}\times U(1)_{\phi }.  \label{symm}
\end{equation}
Here the indices $L$, $S$, and $\phi $ indicate orbital space, spin space
and gauge, respectively.

Clearly, this symmetry group is much richer in structure than
those considered before. In fact, it incorporates the order
parameter symmetries of liquid crystals, $SO(3)_{L}$, as well as
of magnets, $SO(3)_{S}$, and of isotropic superfluids, $U(1)_{\phi
}$,\ all at the same time. This implies that the $broken$
symmetries resulting from (\ref{symm}) will also be much more
intricate than, say, in the case of the ferromagnet, where only a
single degree of freedom is relevant.

One might think that a study of the above mentioned broken symmetries could
be performed much more easily by investigating them separately, i.e. within
the isotropic superfluid, the magnet, the liquid crystal, etc., itself.
However, the combination of several {\em simultaneously} broken continuous
symmetries is much more than just the simple sum of the properties of all
these known systems. Namely, some of the symmetries of superfluid $%
^{3}$He are jointly broken, leading to what is called ``relative''
symmetries. For example, the spin-orbit rotation symmetry, i.e.,
the $SO(3)_{L}\times SO(3)_{S}$-part of (\ref{symm}), may be
simultaneously broken to become $SO(3)_{L+S}$, i.e. a (sub-)group
describing a {\em linear combination} of the individual groups
$SO(3)_{L}$ and $SO(3)_{S}$. This situation is realized in the B
phase of superfluid $^{3}$He (see below). Another example is the
gauge-orbit symmetry, i.e., the $SO(3)_{L}\times U(1)_{\phi
}$-part of (\ref{symm}),
which may be simultaneously broken to a single subgroup $U(1)_{L_{z}+\phi }$%
\cite{Leggett72,Liu78}. This is the case in the A phase of
superfluid $^{3}$He (see below). In both cases a rigid connection
is established between the previously independent degrees of
freedom, leading to long-range order in the condensate only in the
{\em combined} (and not in the individual) degrees of freedom.

\subsection{The structure of the superfluid phases of $^{3}$He}

It is clear that in principle the internal degrees of freedom of a
spin-triplet p-wave state allow for many different Cooper-pair states and
hence superfluid phases. (This is again different from ordinary
superconductivity with $S=0,l=0$ pairing, where only a {\it single} phase is
possible). Of these different states, the one with the lowest energy for
given external parameters will be realized.

\subsubsection{B phase}

In fact, Balian and Werthamer \cite{Balian63} showed, that, within a
conventional ``weak-coupling'' approach, of all possible states there is
precisely one state (the ``BW state'') that has the lowest energy at {\it all%
} temperatures. This state is the one that describes the B phase of
superfluid $^{3}$He. The state originally discussed by these authors is one
in which the orbital angular momentum ${\it {\bf l}}$ and spin ${\it {\bf S}}
$ of a Cooper pair couple to a total angular momentum ${\it {\bf J}}={\it
{\bf l}}+{\it {\bf S}}=0$. This $^{3}P_{0}$ state is, however, only a
special case of a more general one with the same energy (in the absence of
spin-orbit interaction), obtained by an arbitrary rotation of the spin axes
relative to the orbital axes of the Cooper-pair wave function. Such a
rotation may be described mathematically by specifying a rotation axis ${%
\hat{{\bf n}}}$ and a rotation angle ${\theta }$. In the B phase all three
spin substates in (1) occur with equal measure. This state has a rather
surprising property: in spite of the intrinsic anisotropy, the state has an $%
{\it isotropic}$ energy gap. (The energy gap is the amount by which the
system lowers its energy in the condensation process, i.e. it is the minimum
energy required for the excitation of a single particle out of the
condensate.) Therefore the B phase resembles ordinary superconductors in
several ways. On the other hand, even though the energy gap is isotropic,
the B phase is intrinsically anisotropic. This is clearly seen in dynamic
experiments in which the Cooper-pair structure is distorted. For this reason
the B phase is sometimes referred to as ``pseudo-isotropic''. Owing to the
quantum coherence of the superfluid state, the rotation axis ${\hat{{\bf n}}}
$ and angle ${\theta }$ characterizing a Cooper pair in the B phase are
macroscopically defined degrees of freedom, whose variation is physically
measurable.

In the B phase gauge symmetry is completely broken (as in conventional
superconductors or superfluid $^{4}$He) and a linear combination of the
spin-orbit symmetry is broken; the remaining symmetry of the order parameter
of that phase corresponds to the group $SO(3)_{L+S}$.

\subsubsection{A phase}

Since in weak-coupling theory the B phase always has the lowest
energy, an explanation of the existence of the A phase of
superfluid $^{3}$He obviously requires one to go beyond such an
approach and to include ``strong--coupling effects''
\cite{Anderson73,Anderson78}; for a review of a systematic
approach see \cite{Serene83}. In view of the fact that, at
present, microscopic theories are not capable of computing
transition temperatures for $^{3}$He, it is helpful to single out
a particular effect that can explain the stabilization of the A
phase over the B phase at least qualitatively. As demonstrated by
Anderson and Brinkman \cite{Anderson73} such a conceptually simple
effect\ -- a feedback mechanism -- indeed exists: the pair
correlations in the condensed state change the pairing interaction
between the $^{3}$He quasiparticles, the modification depending on
the actual state itself. As a specific mechanism, these authors
considered the role of spin fluctuations and showed that a
stabilization of the state first considered by Anderson and Morel
\cite{Anderson60,Anderson61} is indeed possible. This only happens
at somewhat elevated pressures, since spin fluctuations become
more pronounced only at higher pressures. This ``ABM state'' does
indeed describe the A phase. It has the property that, in contrast
with $^{3}$He-B, its magnetic suspectibility is essentially the
same as that of the normal liquid. This is a clear indication that
in this phase the spin substate with $S_{z}$ = 0, which is the
only one that can be
reduced appreciably by an external magnetic field, is absent. Therefore $%
^{3} $He-A is composed only of $\mid \uparrow \uparrow \rangle $ and $\mid
\downarrow \downarrow \rangle $ Cooper pairs. This implies that the
anisotropy axis of the spin part of the Cooper pair wave function, called $%
\hat{{\bf d}}$, has the same fixed direction in every pair. (More precisely,
$\hat{{\bf d}}$ is the direction along which the total spin of the Cooper
pair vanishes: $\hat{{\bf d}}\cdot {\bf {S}}$ = 0). Likewise, the direction
of the relative orbital angular momentum $\hat{{\bf l}}$ is the same for all
Cooper pairs. Therefore in the A phase the anisotropy axes $\hat{{\bf d}}$
and $\hat{{\bf l}}$ of the Cooper-pair wave function are long-range-ordered,
i.e. are preferred directions in the whole macroscopic sample. This implies
a pronounced anisotropy of this phase in all its properties. In particular,
the value of the energy gap now explicitly depends on the direction in ${\bf
{k}}$ space on the Fermi sphere and takes the form
\begin{equation}
\Delta _{\hat{{\bf k}}}(T)=\Delta _{0}(T)[1-(\hat{{\bf k}}\cdot \hat{{\bf l}}%
)^{2}]^{1/2}.
\end{equation}
Hence the gap vanishes at two points on the Fermi sphere, namely
along $\pm \hat{{\bf l}}$. Because of the existence of an axis
$\hat{{\bf l}}$, this state is also called the ``axial state''.
The existence of nodes implies that in general quasi-particle
excitations may take place at arbitrarily low temperatures.
Therefore, in contrast with $^{3}$He-B or ordinary
superconductors, there is a finite density of states for
excitations with energies below the average gap energy, leading
for example to a specific heat proportional to $T^{3}$ at low
temperatures. The existence of point nodes (''Fermi points'') of
the A phase order parameter allows one to draw detailed analogies
with relativistic quantum field theory
\cite{Volovik92,Volovik00b}, as will be discussed in Section
\ref{connections}.

In the A phase the spin rotation symmetry is broken (as in a
ferromagnet), as well as a linear combination of the gauge-orbit
symmetry. Neglecting an additional discrete $Z_{2}$ symmetry, the
remaining symmetry of the order parameter of that phase therefore
corresponds to the group $U(1)_{S_{z}}\times U(1)_{L_{z}+\phi }$.

\subsubsection{A$_{1}$ phase}

The third experimentally observable superfluid phase of $^{3}$He,
the $A_{1}$ phase, is only stable in the presence of an external
magnetic field. In this phase Cooper pairs are all in a single
spin substate, the $\mid \uparrow \uparrow \rangle $ state,
corresponding to $S_{z}$ = + 1; the components with $\mid \uparrow
\downarrow \rangle $ + $\mid \downarrow \uparrow \rangle $ and
$\mid \downarrow \downarrow \rangle $ states are missing. It is
therefore a ``magnetic'' superfluid -- the first homogeneous
magnetic liquid ever observed in nature. The remaining symmetry of
that phase corresponds to $U(1)_{S_{z}+\phi }\times
U(1)_{L_{z}+\phi }$.

\subsection{Broken symmetries in high-energy physics}

Symmetries and symmetry breaking are of fundamental importance in
high-energy physics. This has to do with the close connection between
symmetries and conserved quantities (Noether's theorem). If, experimentally,
a conserved quantity is identified, the question arises as to what symmetry
is responsible for it. For example, a one-parameter quantity (a charge, say)
involves an underlying $U(1)$ symmetry. One then has to reconstruct the
initial underlying symmetry that by symmetry breaking leads to the
experimentally observable lower symmetries. In this way high-energy physics
may be understood as a sequence of symmetry breakings. We should note,
however, that in high-energy physics the relevant symmetries are usually $%
local$ (i.e. gauge-) symmetries rather than the global symmetries
considered in condensed-matter physics. (This implies an
Anderson-Higgs mechanism rather than Goldstone modes
\cite{Anderson63,Higgs64}). On the other hand, even in high-energy
physics, namely in the theory of strong interactions, there is an
example of a globally broken symmetry that is very similar to
those considered here.

For this let us consider an isodoublet of quarks with only two
flavors, namely an up- and a down-quark ($u$,$d$). Initially there
is a chiral invariance between left-handed ($L$) and right-handed
($R$) states (we neglect the weak interaction):

\begin{equation}
\left(
\begin{array}{c}
u \\
d
\end{array}
\right) _{L},\left(
\begin{array}{c}
u \\
d
\end{array}
\right) _{R}.  \label{quarks}
\end{equation}

The corresponding symmetry, i.e. separate rotations with respect to left and
right, is represented by$\ $%
\begin{equation}
G=SU(2)_{L}\times SU(2)_{R}.  \label{highenergy}
\end{equation}

(For simplicity, we have neglected two additional $U(1)$
symmetries in (\ref {highenergy}): one related to baryon-number
conservation and another connected with an axial symmetry; the
latter is not realized in nature and is believed to be broken by
strong interactions (''the $U(1)$ problem'' \cite{Witten79})).
Experimentally, however, we know that the invariance under the
full symmetry group (\ref{highenergy}) is not observed - a fact
attributed to a broken symmetry. Indeed, one imagines a
condensation between quarks ($u$,$d$) and antiquarks ($\overline{u}$,$%
\overline{d}$) to take place - very similar to Cooper pairing. In this way
the symmetry (\ref{highenergy}) is broken down to a remaining $SU(2)_{L+R}$,
in analogy to the B phase. Here we again encounter a relative broken
symmetry where now the corresponding Goldstone bosons are the three pions.

A different example, now involving gauge symmetries in the theory of
electroweak interactions, is a model due to Pati and Salam \cite{Pati74}, an
extension of the Weinberg-Salam model. Again we consider left/right
isodoublets of $u$, $d$ quarks, and, in addition, a hypercharge. The
left/right symmetries are both described by an $SU(2)$, the latter one by a $%
U(1)$ symmetry. Hence the underlying symmetry is

\begin{equation}
G=SU(2)_{L}\times SU(2)_{R}\times U(1)_{Y/2}
\end{equation}
(note that this group is essentially identical with the symmetry
group (\ref{symm}) describing superfluid $^{3}$He!). On an
experimental level at low energies only the subgroup
$H=U(1)_{L_{z}+Y/2}$, i.e. ordinary QED, is realized. It involves
the $z$ components of the left-handed isospin (the $SU(2)_{R}$ is
assumed to be completely broken) and the hypercharge, whose
generator is the ordinary electric charge $Q=L_{z}+Y/2$. This
implies a broken ''relative'' symmetry involving two symmetries.

\section{Orientational effects on the order parameter}

\label{orient}

For a pair-correlated superfluid, the pairing interaction is the
most important interaction, since it is responsible for the
formation of the condensate itself. Nevertheless, there also exist
other, much weaker, interactions, which may not be important for
the actual transition to the pair-condensed state, but which do
become important if their symmetry differs from the
aforementioned. In particular, they may be able to break remaining
degeneracies.
\newline
\newline
\noindent {\bf The dipole--dipole interaction}: The dipole--dipole
interaction between the nuclear spins of the $^{3}$He atoms leads to a very
weak, spatially strongly anisotropic, coupling. The relevant coupling
constant $g_{D}(T)$ is given by
\begin{equation}
g_{D}(T)\approx \frac{\mu _{0}^{2}}{a^{3}}\left( \frac{\Delta (T)}{E_{F}}%
\right) ^{2}n
\end{equation}
Here $\mu _{0}$ is the nuclear magnetic moment, such that $\mu
_{0}^{2}/a^{3} $ is the average dipole energy of two particles at relative
distance $a$ (the average atomic distance), while the second factor measures
the probability for these two particles to form a Cooper pair and $n$ is the
overall particle density. Since $\mu _{0}^{2}/a^{3}$ corresponds to about $%
10^{-7}$K, this energy is extremely small and the resulting interaction of
quasiparticles at temperatures of the order of $10^{-3}$K might be expected
to be completely swamped by thermal fluctuations. This is indeed true in a
normal system. However, the dipole-dipole interaction implies a spin-orbit
coupling and thereby has a symmetry different from that of the pairing
interaction. In the condensate the symmetries with respect to a rotation in
spin and orbital space are spontaneously broken, leading to long-range order
(for example of ${\bf \hat{d}}$ and ${\bf \hat{l}}$ in the case of $^{3}$%
He-A). Nevertheless, the pairing interaction does not fix the
$relative$ orientation of these preferred directions, leaving a
continuous degeneracy. As pointed out by Leggett
\cite{Leggett73a,Leggett73b,Leggett74,Leggett75}, in this
situation the tiny dipole interaction is able to lift the
degeneracy, namely by choosing that particular relative
orientation of the long-range ordered preferred directions for
which the dipolar energy is minimal. Thereby this interaction
becomes of $macroscopic$ importance. One may also view this effect
as a $permanent$ local magnetic
field of about 3 mT at any point in the superfluid (in a liquid!). In $^{3}$%
He-A the dipolar interaction is minimized by a parallel orientation of ${\bf
\hat{d}}$ and ${\bf \hat{l}}$. \newline
\newline
\noindent {\bf Effect of a magnetic field}: An external magnetic field acts
on the nuclear spins and thereby leads to an orientation of the preferred
direction in spin space. In the case of $^{3}$He-A the orientation energy is
minimal if ${\bf \hat{d}}$ is perpendicular to the field ${\bf H}$, since
(taking into account ${\bf \hat{d}\cdot S}$ = 0) this orientation guarantees
${\bf S}{\Vert }{\bf H}$. \newline
\newline
\noindent {\bf Walls}: Every experiment is performed in a volume of finite
size. Clearly, the walls will have some effect on the liquid inside. In
superfluid $^{3}$He this effect may readily be understood by using a simple
picture. Let us view the Cooper pair as a kind of giant ``molecule'' of two $%
^{3}$He quasiparticles orbiting around each other. For a pair not to bump
into a wall, this rotation will have to take place in a plane parallel to
the wall. In the case of $^{3}$He-A, where the orbital angular momentum $%
{\bf \hat{l}}$ has the same direction in all Cooper pairs (standing
perpendicular on the plane of rotation), this means that ${\bf \hat{l}}$ has
to be oriented perpendicular to the wall. So there exists a strict
orientation of ${\bf \hat{l}}$ caused by the walls \cite{Ambegaokar74}. In
the B phase, with its (pseudo) isotropic order parameter, the orientational
effect is not as pronounced, but there are qualitatively similar boundary
conditions.

\subsection{Textures and defects}

From the above discussion, it is clear that the preferred directions ${\bf
\hat{l}}$ and ${\bf \hat{d}}$ in $^{3}$He-A are in general subject to
different, often competing, orientational effects (for simplicity, we shall
limit our description to $^{3}$He-A). At the same time, the condensate will
oppose any spatial variation of its long-range order. Any ``bending'' of the
order-parameter field will therefore increase the energy, thus giving an
internal stiffness or rigidity to the system. While the orientational
effects might want ${\bf \hat{d}}$ and ${\bf \hat{l}}$ to adjust on the
smallest possible lengthscale, the bending energy wants to keep the
configuration as uniform as possible. Altogether, the competition between
these two opposing effects will lead to a smooth spatial variation of ${\bf
\hat{d}}$ and ${\bf \hat{l}}$ throughout the sample, called a ``texture''.
This nomenclature is borrowed from the physics of liquid crystals, where
similar orientational effects of the preferred directions occur.

The bending energy and all quantitatively important orientational energies
are invariant under the replacement ${\bf \hat{d}\rightarrow -\hat{d},\hat{l}%
\rightarrow -\hat{l}}$. A state where ${\bf \hat{d}}$ and ${\bf \hat{l}}$
are parallel therefore has the same energy as one where ${\bf \hat{d}}$ and $%
{\bf \hat{l}}$ are antiparallel. This leads to two different,
degenerate ground states. There is then the possibility that in
one part of the sample the system is in one ground state and in
the other in a different ground state. Where the two
configurations meet they form a planar ``defect'' in the texture,
called a ``domain wall'' \cite{Maki77}. This is in close analogy
to the situation in a ferromagnet composed of domains with
different orientations of the magnetization. Domain walls are
spatially localized and are quite stable against external
perturbations. In fact, their stability is
guaranteed by the specific nature of the order-parameter structure of $^{3}$%
He-A. Mathematically, this structure may be analysed according to its
topological properties; for reviews see the articles by Mermin \cite
{Mermin79} and Mineev \cite{Mineev80}. The stability of a domain wall can
then be traced back to the existence of a conserved ``topological charge''.
Using the same mathematical approach, one can show that the order-parameter
fields of the superfluid phases of $^{3}$He not only allow for planar
defects but also for point and line defects, called ``monopoles'' and
``vortices'', respectively. Defects can be ``non-singular'' or ``singular''
, depending on whether the core of the defect remains superfluid or whether
it is forced to become normal liquid. The concept of vortices is, of course,
well-known from superfluid $^{4}$He. However, since the order-parameter
structure of superfluid $^{3}$He is so much richer than that of superfluid $%
^{4}$He, there exist a wide variety of different vortices in these
phases. Their detailed structure has been the subject of intensive
investigation, in particular in the context of experiments on
rotating superfluid $^{3}$He, where they play a central role
\cite{Hakonen87,Salomaa87}.

\section{Relation to other fields}

\label{connections}

Why spend so much effort on sorting out the strange behavior of
states of matter only found at temperatures so low that they are
even outside the reach of most low-temperature laboratories?
Partly, of course,
``because it's there'', and because -- like any other system -- superfluid $%
^{3}$He deserves to be studied in its own right. However, what is even more
important is that superfluid $^{3}$He is a model system that exemplifies
many of the concepts of modern theoretical physics and, as such, has given
us, and will further provide us, with new insights into the functioning of
quantum-mechanical many-body systems close to their ground state.

\noindent {\bf Quantum amplifier:} The macroscopic coherence of superfluid $%
^{3}$He may serve as a ''quantum amplifier'' of weak and ultraweak
interaction effects. One example (see Section \ref{orient}) is the
dipole-dipole interaction between the Helium nuclei. Even more
spectacular is the prediction by Leggett \cite{Leggett77a} of a
permanent electric dipole moment of the Cooper pairs in the B
phase due to the amplification of the parity-violating part of the
electron-proton interaction (''weak neutral currents'') mediated
by the exchange of $Z^{0}$ bosons; for a detailed discussion see
\cite{Vollhardt90}. The permanent parity-violating dipole moment
in the B phase is quite small but is not hopelessly outside the
experimentally accessible range. The weak interaction between
electrons and hadrons also has a $T$- (i.e. $CP$-) violating part.
A measurement of the electric dipole moment of the B phase would
then allow one to determine the strength of the $T$-odd
interaction \cite{Khriplovich82}.

\noindent {\bf Unconventional superconductors:} Anisotropic
superconducting states are particularly exciting. \ It is now well
established that superconductivity in the high-$T_{c}$ cuprates
and, at least in some cases, in the so-called ``heavy-fermion''
systems, is due to the formation of anisotropic pairs with d-wave
symmetry \cite{Cox95}. Many of the concepts and ideas developed
for superfluid $^{3}$He have been adapted to these systems.

\noindent {\bf Neutron stars: }There are several other physical
systems for which the ideas developed in the context of superfluid
$^{3}$He are relevant. One or them is an anisotropic superfluid
system that already exists in nature but is not accessible for
laboratory experiments: this is the nuclear matter forming the
cores of neutron stars. There the pairing of neutrons has been
calculated to be of p-wave symmetry. Because of the strong
spin-orbit nuclear force, the total angular momentum of the Cooper
pairs is predicted to be $J$ = 2 \cite{Sauls82,Pines85}.

\noindent {\bf Cosmology and early universe:} Recently superfluid $^{3}$He
has been used as a test system for the creation of ``cosmic strings'' in the
early stages of the universe. According to Kibble \cite{Kibble76} and Zurek
\cite{Zurek85} the observed inhomogeneity of matter in the universe may be
understood as the result of the creation of defects generated by a rapid
cooling through second-order phase transitions, which led to the present
symmetry-broken state of the universe. In two different experiments with
superfluid $^{3}$He, performed at the low-temperature laboratories in
Grenoble \cite{Bäuerle96} and Helsinki \cite{Ruutu96}, a nuclear reaction in
the superfluid, induced by neutron radiation, caused a local heating of the
liquid into the normal state. During the subsequent, rapid cooling back into
the superfluid state the creation of a vortex tangle was observed. The
experimentally determined density of this defect state was found to be
consistent with Zurek's estimate and thus gives important support to this
cosmological model; see also \cite{Eltsov00}. Furthermore, a recent
experimental verification of momentogenesis in $^{3}$He-A by Bevan et al.
\cite{Bevan97} was found to support current ideas on cosmological
baryogenesis. (Baryogenesis during phase transitions in the early universe
is believed to be responsible for the observed excess of matter over
antimatter). In view of these exciting new developments it may become
possible in the future to model and study cosmological problems in the
low-temperature laboratory in much more detail.

\noindent {\bf Particle physics and gravity:} As discussed in
Section \ref {brokensymm}, the key to understanding superfluid
$^{3}$He is ``spontaneously broken symmetry''. Volovik showed that
the existence of point nodes (''Fermi points'') in the A phase
order parameter allows one to draw detailed analogies with
relativistic quantum field theory \cite
{Volovik87,Volovik92,Volovik99,Volovik00a,Volovik00b}. Thereby
fundamental connections with particle physics exist which derive
from the interpretation of the order-parameter field as a quantum
field with a rich group structure. The collective modes of the
order parameter as well as the localized topological defects in a
given ground-state configuration are the particles of this quantum
field theory. Various anomalies (e.g. the axial anomaly) known
from particle physics can be identified in the $^{3}$He model
system, and it appears that fundamental insight gained from the
experimental study of superfluid $^{3}$He will be useful in
particle theory. Indeed, both the A phase of superfluid $^{3}$He
and the electroweak vacuum belong to the class of gapless Fermi
systems with topologically stable point nodes \cite{Volovik92}.
The fermionic (quasi-)particles in this class are chiral, and the
collective bosonic modes are the effective gauge and gravitational
fields \cite{Volovik99,Volovik00b}. This provides a direct link
between the dynamics of quantum liquids and gravity
\cite{Volovik00a}. In particular, in the limit of very low
energies these condensed matter systems reproduce all symmetries
known today in high energy physics, e.g. Lorentz invariance, gauge
invariance and general covariance.

Clearly, the unique richness of the structure of the superfluid
phases of $^{3}$He continues to lead to new aspects whose
investigation provides unexpected insights.

\label{Refs}

%\newpage
\bigskip

FIGURE CAPTIONS\\ \\{\em Fig. 1:} Pressure vs. temperature phase
diagram for (a) $^3$He (logarithmic temperature scale); (b) $^4$He
(linear temperature scale).

\end{document}